\documentclass[epj,referee]{svjour}
\usepackage{graphics}
\begin{document}
\title{Proposal for a Raman X-ray Free Electron Laser}
\author{Ph. Balcou
}                     
%
%
\institute{Universit\'e de Bordeaux, Centre Lasers Intenses et Applications (CELIA), CNRS, CEA, 351 Cours de la lib\'eration, F--33405 Talence, FRANCE}
\date{Received: date / Revised version: date}
\date{version 29 mai 2009}
%
\abstract{
A scheme for an X-ray free electron laser is proposed, based on a Raman process occurring during the interaction between a moderately relativistic bunch of free electrons, and twin intense short pulse lasers interfering to form a transverse standing wave along the electron trajectories. In the high intensity regime of the Kapitza-Dirac effect, the laser ponderomotive potential forces the electrons into a lateral oscillatory motion, resulting in a Raman scattering process. I show how a parametric process is triggered, resulting in the amplification of the Stokes component of the Raman-scattered photons. Experimental operating parameters and implementations, based both on LINAC and Laser Wakefield  Acceleration techniques, are discussed.
} 
\maketitle
\section{Introduction}
\label{intro}

Obtaining a laser effect in the extreme Ultraviolet and X-ray ranges has long been a major objective in laser science. The first proposals and attempts started in the late sixties and early seventies with the first contributions of Duguay and Rentzepis \cite{DuguayRentzepis}, and of Jaegl\'e \cite{jaegle1}. After almost thirty years of research coupling the physics of lasers and of plasmas used as active media, numerous lasing lines have been demonstrated and brought to saturation in the extreme ultraviolet and very soft X-ray ranges.
In parallel, the progress of ultrashort pulse intense lasers have led to other scenarios  : high harmonic generation is now a well established method to use extreme non-linear optics in order to create laser-like radiation in the XUV spectral range. Both high harmonic and soft X-ray lasers from laser/plasma interactions are reviewed in a recent textbook \cite{jaegle}.

However, these processes are usually limited to photon energies of few hundreds of eV; the conversion efficiency from laser  to XUV pulses drops at higher photon energies, which limits severely the applicability of laser-plasma X-ray lasers and high harmonic generation, in the soft to hard X-ray ranges.

Two main paths have been followed in recent years to obtain intense X-ray pulses : X-ray free electron lasers, and incoherent X-ray emission during the interactions between intense lasers, and matter under various phases : solids, clusters, relativistic free electrons.

After the pioneering proposals and first developments of Free Electron Lasers \cite{madey,deacon}, it was proposed to extend the concept to the extreme ultraviolet and soft X-ray ranges \cite{murphy,pellegrini}. This resulted in large scale X-ray free electron projects in the US (LCLS), Japan (SCSS) and Europe (FLASH and TESLA X-FEL) \cite{ayvazyan}. A huge potential of new applications is expected in many sciences, from physics to biology; the cost and size of these projects are obvious limiting factors to a widespread use of X-ray free electron lasers.

\textit{A contrario}, facilities to generate X-ray pulses  during the interaction of intense lasers and matter are compact, less costly, but yield mostly incoherent light, with brightnesses smaller by several decades. Of particular interest here is the process of Thomson (or inverse Compton) scattering of laser light, in which photons from a high power laser impinge on a bunch of moderately relativistic electrons , and scatter with an important Doppler shift of $4\gamma^2$, $\gamma$ denoting the Lorentz factor, thus appearing in the laboratory frame as X-ray photons, collimated in a small angle \cite{sprangle92}.
The main advantage of this laser/free electron interaction process is the compacity of the setup : scattering real laser photons, whose wavelength is in the micrometer range, allows to reach X-ray wavelengths with Lorentz factors of typically $10^2$, whereas a Lorentz factor of $10^4$ or more  is required to scatter virtual photons of an undulator, with a period of a few centimeters.
Electron energies up to 50 MeV only are therefore required with laser scattering, which can be obtained with a small linac of only few meters, instead of few kilometers necessary to reach multi-GeV energies.

Being able to combine both schemes in order to blend their attractive features: compacity of laser scattering, and coherent amplification of a X-ray free-electron laser, would be extremely appealing.
As an attempt in this direction, many authors have emphasized that the action of a laser field, propagating in the opposite direction to the relativistic electrons, is extremely similar to that of the magnetic field within an undulator. A laser-undulator free electron laser has therefore been repeatedly proposed \cite{sprangle79,dobiasch,gea,gallardo}. However, the strength parameter $K$ of the laser undulator remains usually very small with most conceivable laser parameters.
The gain per oscillation period is then severely reduced with respect to normal undulators, which implies to force the electrons to wiggle a very large number of times N during the amplification. Since the level of mono-energeticity of the electron bunch has to be smaller than $1/2N$ for the Compton free electron laser effect to be effective \cite{saldin}, this scheme would require a quality of mono-energeticity  beyond the present state of the art, as well as a remarkably flat intensity profile of the laser pulse, both temporally and spatially.
Numerical studies were performed by Bacci {\textit et al.} \cite{Bacci}, that predict a coherent enhancement of extreme UV radiation, considering however a remarkable mono-energeticity of $10^{-4}$, and extremely ambitious laser parameters.

In a variant of this scheme, several contributions predict novel phenomena at the onset of quantum effects, expected again only for outstanding qualities of mono-energeticity \cite{bonifacio1,bonifacio2,avetissian}.
Finally, laser wakefield acceleration of electrons is increasingly considered as a potential compact substitute of  conventional accelerator technology, at least for extreme UV free electron lasers \cite{gruner,Nakajima}. A scheme coupling laser acceleration of electrons , and a laser undulator,  can open the way to an all-optical Xray free electron laser \cite{Petrillo}.
However, all these schemes require very stringent parameters of mono-energeticity and emittance of electron bunches, and seem  extremely challenging in view of present day electron and laser technologies.

We explore here  an alternative opportunity to create a compact X-ray FEL, by coupling the physics of free electron lasers, of laser-plasma XUV lasers, and of extreme non-linear optics.
By creating artificially a quasi-internal degree of freedom to relativistic free electrons dressed by intense optical lasers, a  non-linear Raman scattering process might be switched, leading to exponential amplification of X-ray light.
A laser-like beam could then be envisioned, starting either through a SASE process (Self-Amplified Spontaneous Emission), or through the injection of a low intensity, soft X-rays beam from high harmonic generation \cite{zeitoun,lambert}.

The setup considered is first depicted, and the electron dynamics described; this allows to unravel the characteristic emission frequencies of the Raman lines. In sec. \ref{sec:amp}, the amplification process is modeled analytically, resulting in the calculation of the gain coefficient.
Finally, the prospects for an experimental test are discussed, in view of the present state of the art in laser and electron accelerator technologies. A survey of the main relevant parameters with conventional or laser wakefield acceleration systems is presented, along with order-of-magnitude estimates of the laser specifications required to achieve lasing in the X-ray range.

\section{Principles of a Raman X-ray FEL}
\label{sec:present}

\subsection{Interaction geometry}
\label{ssec:intergeom}

As schematized in Fig. 1, let us consider the interaction between :

i) a bunch of free electrons, as issued either from a linear accelerator or a small storage ring, with a kinetic energy in the range from 10 to 50 MeV, and hence a Lorentz factor from 20 to 100; the propagation axis of the electrons is taken as the conventional $z$-axis. This element is typical of Thomson (inverse Compton) scattering experiments, or of free electron laser, except for the use of smaller electron kinetic energies;

ii) a femtosecond or picosecond intense laser system, whose beam is split into two strictly identical parts. The twin beams are made to counter-propagate with respect to one another,   along the $x$-axis perpendicular to the electron direction. The polarization vector of the twin beams will be chosen as linear along the $y$ axis (vertical linear polarization, if $xz$ is considered as an horizontal plane). In this configuration, the magnetic field of the twin lasers is along the $z$ axis.

\begin{figure}
\resizebox{1.1\columnwidth}{!}{%
  \includegraphics{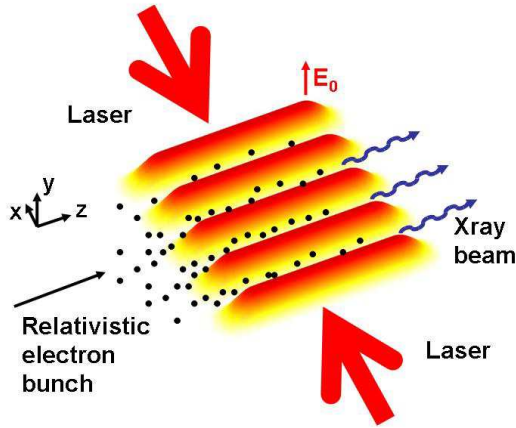}
}
\caption{Proposed configuration for a Raman X-ray laser. The color code from yellow to red indicates the height of the ponderomotive potential due to the laser standing wave.}
\label{fig:1}       
\end{figure}

Both beams are focused along a line, in order to overlap in space and time over the electron path. This superposition of the laser beams along the $z$ axis results in the formation of a standing wave along $x$.
The beam intensity along the focal line will have to be controlled to be as constant as possible, after a beam ramp-up segment, and will be given a spatial profile as flat as possible. Such constraints are similar to those encountered for optical parametric chirped pulse amplification systems, and can be fulfilled by means of
 high quality optical elements and spatial phase control devices, available with present day technologies. This may also ensure that the positions along $x$ of the nodes of the standing wave are constant along the propagation direction $z$.

\begin{figure}
\resizebox{1.1\columnwidth}{!}{%
  \includegraphics{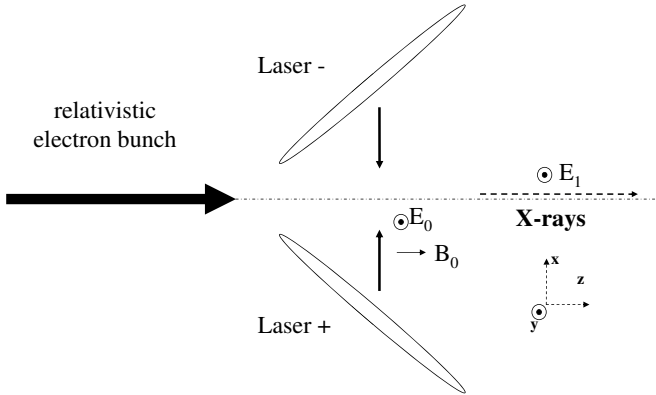}
}
\caption{Configuration of the two counter-propagating laser inhomogeneous waves. The energy front is oriented at 45$^\circ$ from the phase fronts, resulting in the advance of the interference region at velocity $c$ along the $z$ axis.}
\label{FigTravWave}       
\end{figure}

 An important point is to synchronize the advance of the electrons, and the illumination  by the twin transverse laser beams. Indeed, most studies of laser-plasma soft X-ray lasers \cite{klisnick} display a similar configuration, in which a transverse high intensity laser impinges at $90^\circ$ onto a solid surface, thus creating an optically active plasma. In most cases, the duration of the population inversion at each point within the plasma is well below the traversal time of the photons in the amplification region; as a result, the transverse illumination by the laser has to be made to follow the displacement of the X-ray photons along the target. This is achieved thanks to a special optical geometry, in which the energy front of the illuminating laser is decoupled from its phase fronts, by means of diffractive elements \cite{bor,chanteloup}.  In this "inhomogeneous wave" geometry (also sometimes referred to as "traveling wave" geometry), the transverse laser should ideally have an energy front oriented at 45$^{\circ}$ from the phase fronts, yielding a displacement of the illumination area at exactly the speed of light. Various variants of the optical implementation of the traveling wave are being considered, with an accuracy at the femtosecond level, in order to explore X-ray laser schemes based on  innershell pumping \cite{sebban}.
 We propose to use such an inhomogeneous wave geometry, in which the inhomogeneous traveling wave is  split into two beams, somehow alike the configuration proposed by Pretzler \textit{et al.} \cite{pretzler} for an inverted field auto-correlator. The twin beams are subsequently focused along a line in a counter-propagating configuration, as shown in Fig. \ref{FigTravWave}. The optical implementation may require to control precisely the angle between the phase and energy front, resulting in a fine tuning of the advance velocity of the superposition region of the twin beams.

At moderately high laser intensities, the electrons in the bunch will then interact with the standing wave in a non-linear way, as explained now.

\subsection{High intensity relativistic Kapitza-Dirac effect : numerical simulation}
\label{sec:KapitzaDirac}

Kapitza and Dirac \cite{kapitza} have shown that electrons interacting with a light standing wave can diffract from this light lattice -- thus undergoing the reverse process of light diffraction on a matter density grating. In the low intensity limit, the interaction with the light is a small perturbation to the electron free motion, that induces a momentum transfer of $\pm 2\hbar \mathbf{k}$, where $\mathbf k$ is the wavevector of either beam forming the standing wave \cite{freimund}.
Conversely, at high intensities of the order of $10^{13}$ W/cm$^2$ or more for near infra-red lasers,
the electron dynamics is modified considerably by the action of the light lattice.
Free electrons interacting with a spatially non uniform laser field are indeed submitted to a significant ponderomotive force, ie, a drift force tending to expel the electrons from the regions of highest intensity \cite{kibble68}. The general expression of the ponderomotive force $F_p$ is :
\begin{equation}
\mathbf{F}_p = -\nabla \frac{e^2E^2}{4m\omega_0^2},
\nonumber
\end{equation}
where $-e$ is the electron charge, $E$ the local electric field, $m$ the electron rest mass, and $\omega_0$ the laser angular frequency.
In this case, non-relativistic electrons injected into the standing wave will feel a ponderomotive force deriving from a spatially oscillating potential :
\begin{equation}
V_p =  \frac{e^2E_0^2}{m\omega_0^2} \sin ^2(k_0 x),
\label{FpondStandWave}
\end{equation}
with $k_0=\omega_0 /c$. If the electron transverse kinetic energy is smaller than the maximum of $V_p$, it will  be trapped within the ponderomotive potential well.
In the opposite case, the electron will succeed in going through the light lattice, with a momentum transfer up to several thousands $\hbar \mathbf{k}$ or more.

Bucksbaum, Schumacher, and Bashkansky have studied experimentally the Kapitza-Dirac effect in the high intensity regime, using Above-Threshold Ionization as the source of electrons within the standing wave \cite{bucksbaum}. Giant momentum transfers were indeed observed; importantly, this study concluded on the validity of a classical description of the electron motion in the high intensity regime.

The major difference between the  experiment of Bucksbaum \textit{et al.}, and our proposed X-ray laser scheme, is related to the relativistic velocity of the injected electrons -- a situation not considered so far. As a first step to explore the electron dynamics, we first present the results of a full numerical integration of the electron trajectory.

\begin{figure}
\resizebox{\columnwidth}{!}{%
  \includegraphics{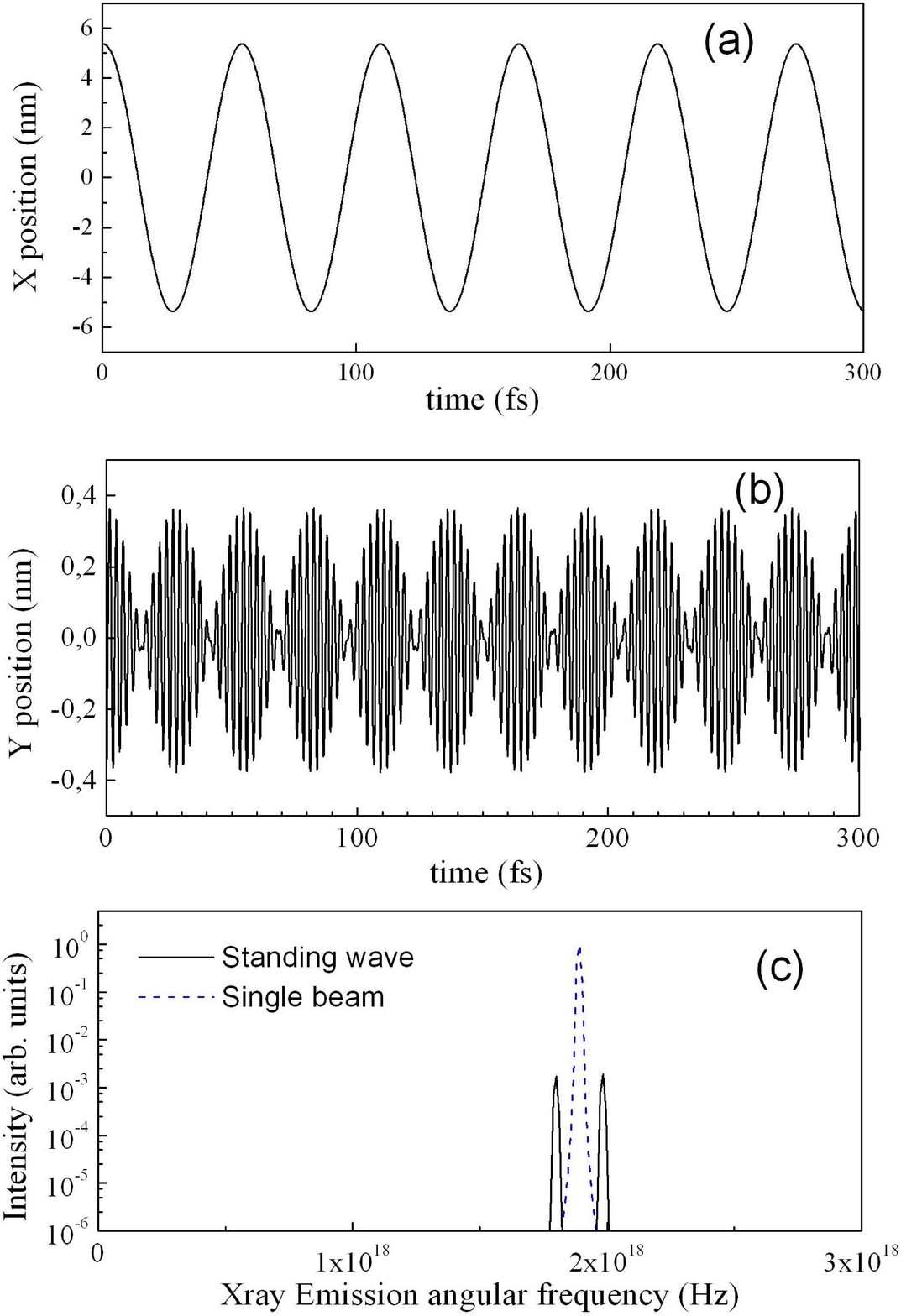}
}
\caption{(a) Transverse motion of a relativistic electron trapped laterally in a laser standing wave. (b) Corresponding laser-induced oscillation. (c) Emission spectrum for an electron in a standing wave (solid line) and in a normal single side illumination (90$^{\circ}$ Thomson scattering), blue dashed line).}
\label{fig:3}       
\end{figure}

As a model case, let us consider a 10 MeV electron ($\gamma = 20$), with a small initial transverse velocity, and embedded in the standing wave. We calculate the electron motion using the exact equations of special relativity, and considering both the electric and magnetic fields of the incident laser waves \cite{RefB_shiozawa} :
\begin{eqnarray*}
 \lefteqn{ \frac{d}{dt}\left[\gamma mc^2\right]  = \mathbf{v}.\mathbf{E} }\\
&& \frac{d\mathbf{v}}{dt} = \frac{q}{\gamma m} \left[\mathbf{E}-\frac{\mathbf{v}}{c^2}(\mathbf{v}.\mathbf{E})+\mathbf{v \times B}\right]
\label{RelatEqu}
\end{eqnarray*}

where all dynamical variables are considered in the laboratory frame.
The laser parameters considered are those of a Titanium-Sapphire laser, with a wavelength of 800 nm, and an intensity per beam of $10^{18}$ W/cm$^2$.  The electron initial transverse velocity along $x$ is 6. 10$^5 m.s^{-1}$.
In Fig. 3(a), the electron is seen to wiggle along $x$ around the minimum line of the ponderomotive potential, with a period of 55 fs in this specific case. This period is not only longer than the laser period $T_0 = 2.5 fs$, and also much longer than the oscillation period of 2.7 fs expected from the non-relativistic potential function (\ref{FpondStandWave}) (see Eq. \ref{decRaman} below).
One should also notice that the electron motion along $x$ is perfectly smooth, even within the time span of the laser  cycle $T_0$ -- the ponderomotive potential can therefore be considered as a tool to model the electron dynamics, even on a time scale smaller than $T_0$.
Fig. 3(b) shows how the slow wiggling along $x$ modifies the laser-induced oscillation, which appears now modulated at twice the wiggling frequency.
Finally, Fig. 3(c) displays the spectrum of the light scattered in the $+z$ direction (solid line), calculated as the squared modulus of the Fourier transform of the acceleration along $y$, and taking into account the Doppler shift. The dashed line shows for comparison the light spectrum calculated for the same electron initial conditions, but assuming one of the twin beams to be suppressed.
The Doppler-shifted emission line, characteristic of 90$^{\circ}$ Thomson scattering, is seen to be split into two Raman components, with an important drop in emission intensity because of the electron trapping close to the potential minimum.

An important issue is how relativistic electrons may be injected into the standing wave. Fig. 4 shows few test cases of electrons, chosen at random in an electron bunch, whose normalized emittance is 1 mm.mrad, focused onto a spot of 50 $\mu m$ radius rms. The standing wave is assumed to start with a 3 $mm$ long ramp, corresponding to 10 $ps$, with  sinus-square intensity profile, followed by a plateau of constant intensity. In a first step, the electron motions are hardly affected by the standing wave; as the latter increases further, the electrons are seen to get trapped in one of the potential wells, with a gradually decreasing excursion from the minimum until the end of the ramp.
The light lattice then acts as a duct, able to confine and guide the electrons up to the end of the illuminated area. In this simple calculation, we do not taken into account any back action of the light field emitted by the wiggling electrons on their trajectories; their oscillations along $x$ remain therefore purely randomly phased up to the end of the interaction region.
\begin{figure}
\resizebox{1.1\columnwidth}{!}{%
  \includegraphics{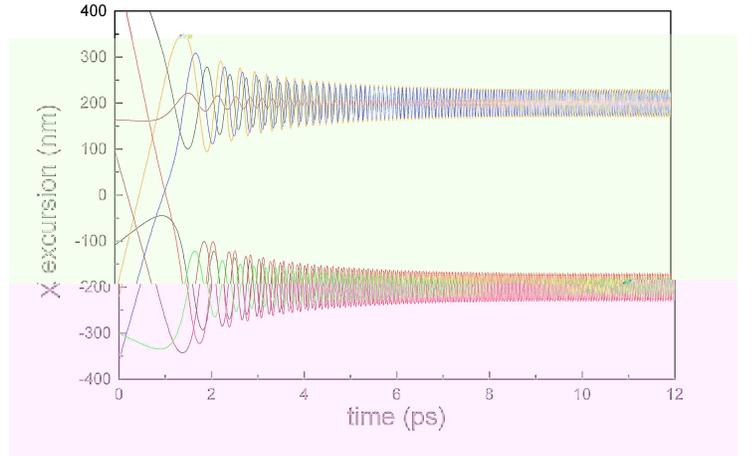}
}
\caption{Test cases of electron injection into the light lattice. Bunch parameters : $\epsilon_N = 1 mm.mrad$, $\gamma = 20$, spot size rms $\sigma _x=50 \mu m$. The standing wave is gradually switched on with a $10 ps$ ramp. The data for figure 3 are taken from one of these trajectories, between 10 and 12 ps.}
\label{fig:4}       
\end{figure}

\subsection{Collective electron motion under X-ray irradiation}
\label{ssec:CollMotion}

We now examine how electrons injected into the light lattice, may be coupled to an external X-ray field, whose frequency corresponds to one of the Raman modes displayed in Fig. 3(c).
We therefore add the possibility to take into account an additional electromagnetic field $E_1(z,t)=E_1^0 \cos(\omega_1 (t-z/c))$, where $\omega_1$ corresponds to the Stokes mode. The X-ray electric and magnetic fields are simply added to the laser fields in the computation of electron motions. We wish to investigate how this X-ray field may modify the distribution in space of the electrons close to the bottom of the potential wells, at a given time.

We consider an initial ensemble of macro-particles, first injected into the light lattice with the same parameters as in Fig. 4, and follow the electrons in time throughout the ramp and the interaction regions.  For the sake of simplicity, we will switch on the X-ray field at the end of the ramp region, and keep it constant up to the end of the interaction region.
In the current example, we restrict the calculation to a slice of phase-space for the initial injection, ie, we consider only electrons initially close to the axis of the potential well, to within $\lambda_0/15$. Taking all electrons at that stage, including the eccentric ones with large amplitude and reduced frequency oscillations, would indeed blur the final figure.

Fig. 5(a) displays the final space distribution (z,x) of an ensemble of 1000 such electrons, with a X-ray field amplitude $E_1^0=10^{10} V/m$, and an interaction region of $75 \mu m$. The region of interest is taken here to have a width of half a laser wavelength, which is the period of the light  lattice, and a length of two X-ray wavelengths $2. (2\pi c /\omega_1)$.
Fig. 5(b) shows the distribution of electrons with identical initial conditions following the ramp, but subjected to  a X-ray field amplitude $E_1^0=10^{10} V/m$ within the interaction region.

\begin{figure}
\resizebox{1.1\columnwidth}{!}{%
  \includegraphics{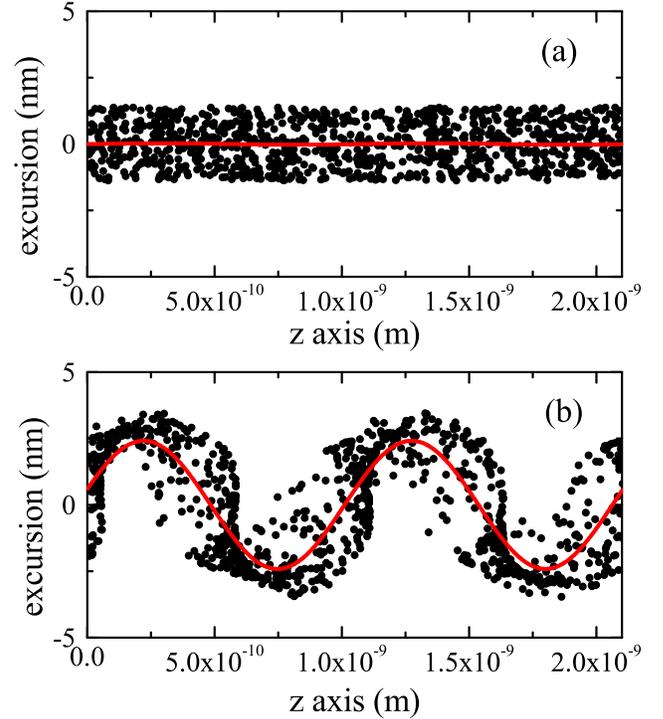}
}
\caption{(a) : space distribution of electrons injected into the light lattice in the conditions of Fig. 4, with no X-ray field. (b) : space distribution after the interaction with a $y$-polarized X-ray field along $z$, at the Stokes frequency. Red lines show least-square fits to sine functions.}
\label{fig:5disp}       
\end{figure}

While each electron oscillates in the light potential well, the random character of the injection into the light lattice results in an evenly distributed electron distribution in Fig. 5(a). On the contrary, one notes easily an overall oscillation of the centroid of electron lateral positions in Fig. 5(b), with the period of the X-ray wavelength along $z$.
The red line is a least-square fit a a sine function to the electron distribution; this gives an intuitive notion of a collective transverse displacement function.
While the detailed process will be unraveled below, it is clear at this stage that the beating between the Doppler-shifted laser frequency, and the Stokes X-ray frequency,  is bound to induce a resonant excitation at the Raman frequency, resulting in this collective behaviour. The same calculation at the anti-Stokes frequency gives absolutely similar distributions.

These numerical results will now allow us to propose an analytical modeling of the electron dynamics, based on a ponderomotive potential approach, and that will consider that electrons remain confined close to the bottom of the potential wells.

\section{Analytical description of  single electron dynamics}
\label{sec:singleEdyn}

\subsection{Analytical description of the single electron dynamics in the light lattice}
\label{ssec:EdynLightLattice}

We model the motion of electrons, moving along the $+z$ direction, and injected into the superposition of two transverse counter-propagating lasers :
one beam in the $+x$ direction, $\mathbf{E}_0^+(x,t)=E_0 \sin(k_0x-\omega_0 t) \mathbf{e}_y$, and an identical beam propagating in the $-x$ direction,  $E_0^-(x,t)=E_0 \sin(k_0x+\omega_0 t)\mathbf{e}_y$.
 $E_0$ is the real-valued electric field associated to each of the twin beams, $k_0$ and $\omega_0$ the laser wavevector and angular frequency respectively, and $\mathbf{e}_y$ (resp. $\mathbf{e}_x$, $\mathbf{e}_z$) will denote the unit polarization vector in the $y$ (resp. $x$,$z$) direction.
These twin laser beams interfere to form a standing wave, described as :
\begin{eqnarray}
\mathbf{E}_0(x,t) &= 2 E_0 \sin (k_0 x) \cos(\omega_0 t)\mathbf{e}_y \\
\mathbf{B}_0(x,t) &= -2 \frac{E_0}{c} \cos (k_0 x) \sin(\omega_0 t)\mathbf{e}_z,
\end{eqnarray}

We assume that the standing wave is switched on adiabatically along $z$ (gradual build-up of the laser intensity along the electron trajectory), and that the transverse kinetic energy of the electron is small with respect of the maximum of the ponderomotive potential $V_p$. Then each electron undergoes an harmonic oscillatory motion close to the bottom lines of $V_p$, with an effective potential given to first order by :
\begin{equation}
V_p^0 =  \frac{e^2E_0^2}{mc^2} x^2
\end{equation}
where for simplicity we have considered small displacements around the minimum potential line $x=0$.
In this harmonic potential well, a non-relativistic electron oscillates with a frequency $\Omega'$:
\begin{equation}
\Omega '=\frac{\sqrt{2}eE_0}{mc}
\label{decRamanNonRel}
\end{equation}
Surprisingly, this oscillation frequency is independent from the laser frequency, but varies as the square root of laser intensity.

Let us consider now a relativistic electron, of velocity $v$ (Lorentz factor $\gamma = (1-v/c)^{-1/2}\gg 1$), as issued from a linear accelerator.
Due to the relativistic mass increase in the laboratory frame, the ponderomotive potential becomes :
\begin{equation}
V_p =  \frac{e^2E_0^2}{\gamma m\omega_0^2} \sin ^2(k_0 x),
\label{VpondStandWaveRelat}
\end{equation}
and the transverse equation of motion close to the bottom of the potential well is :
\begin{equation}
\gamma m x^{..}+ \frac{2 e^2 E_0^2}{\gamma m c^2} x = 0
\end{equation},
which yields an oscillation frequency :
\begin{equation}
\Omega=\frac{\sqrt{2}eE_0}{\gamma mc},
\label{decRaman}
\end{equation}
in excellent agreement with the numerical values obtained from the exact numerical calculation of section \ref{sec:KapitzaDirac}.
One alternative way to obtain the same expression is to transform the standing wave to the electron rest frame, evaluate the oscillation frequency (\ref{decRamanNonRel}) in the ponderomotive potential, and transform the frequency back to the laboratory frame, thus yielding the same expression (\ref{decRaman}).

The position of an arbitrary electron $i$ can hence be described in the laboratory frame as :
\begin{eqnarray}
&z_i^0(t) &= z_i^0(0)+vt \nonumber\\
&x_i^0(t) &= \Delta X_i \cos(\Omega t - \Psi_i) \label{freeXmotion}\\
&y_i^0(t) &= y_i^D(t) + \frac{eE_0k_0\Delta X_i}{\gamma m} \sum_{\epsilon =\pm 1} \frac{\cos((\omega_0+\epsilon\Omega) t - \epsilon\Psi_i)}{(\omega_0+\epsilon\Omega)^2}\nonumber
\end{eqnarray}
where the initial position $z_i^0(t)$, the  vertical drift $y_i^D(t)$, the excursion  $\Delta X_i$ and phase $\Psi_i$ of the free oscillation along $x$,  all result from the initial injection conditions of the electron in the standing wave.
We recover in this simple model that the wiggling in the $y$ direction is split into two Raman shifted lines, of frequencies $\omega_0 \pm \Omega$. The electron oscillation induces light scattering, which, along the electron direction, occurs at frequencies :
\begin{equation}
\omega_1=\frac{\omega_0+\epsilon \Omega}{1-v/c},
\label{FreqOm1}
\end{equation}
where $\epsilon =+1$ corresponds to the anti-Stokes Raman component, and $\epsilon =-1$ to the Stokes component. The $1-v/c$ factor results from the Doppler shift, and corresponds to a frequency up-shift of $2\gamma^2$ in the highly relativistic limit. These analytical values agree again with those displayed on Fig. 3(c).

\subsection{Single electron coupling to a Raman scattered wave}
\label{ssec:spa}
Let us now consider the coupling between the single electron dynamics in the standing wave, and a Raman scattered X-ray wave $E_1$ propagating along $z$:
\begin{equation}
\mathbf{E}_1(z,t)= E_1 \cos (\omega_1t - k_1 z) \mathbf{e}_y
\end{equation}
where $k_1$ is the wave-vector along $+z$ corresponding to the angular frequency  $\omega_1$ given by Eq. \ref{FreqOm1}.
This field is assumed to be polarized along $y$, since it results from the scattering of the  $y$-polarized laser beams.
The magnetic field $\mathbf{B}_1$ of this X-ray wave is therefore directed along $x$.
Each electron will see its motion modified by the coupling of the laser standing wave, and of the X-ray wave, via  Lorentz forces.
Two  terms can be distinguished : $E_1$ induces a small amplitude wiggling around $y$ that couples to the large magnetic field of the standing wave along $z$, resulting in a Lorentz force along $x$; and $E_0$ induces a large wiggling along $y$, that couples to the initially small magnetic field of the X-ray wave, resulting in a second Lorentz force, directed along $z$.
It can easily be shown that these two terms have exactly the same magnitude; however, the latter is obviously non resonant, whereas we will show hereunder that the former induces a resonant oscillation of the electron captured within the ponderomotive potential wells.
The electric force experienced by electron $i$ due to the X-ray field $E_1$ is :
\begin{eqnarray}
{\mathbf F}_1(t) &= -e E_1 \cos \left(\omega_1t - k_1(z_i^0+vt)\right)\mathbf{e}_y \nonumber\\
&= -e E_1 \cos \left( (\omega_0 + \epsilon \Omega)t - \Phi_i)\right) \mathbf{e}_y\label{F1part},
\end{eqnarray}
where $\Phi _i = k_1 z_i^0$.
The resulting wiggling velocity
\begin{equation}
{\mathbf v}_1 (t)=\frac{-eE_1}{\gamma m(\omega_0 + \epsilon \Omega)} \sin \left( (\omega_0 + \epsilon \Omega)t - \Phi_i)\right)\mathbf{e}_y
\end{equation}
couples to the laser magnetic field to yield a transverse Lorentz force :
\begin{equation}
{\mathbf F}_L (t)=\frac{-eE_1 E_0}{\gamma mc(\omega_0 + \epsilon \Omega)} \cos (\epsilon \Omega t - \Phi_i) \mathbf{e}_x
\end{equation}
where we have neglected a rapidly oscillating term at angular frequency $2\omega_0$.
The $x$-motion follows therefore the following equation :
\begin{equation}
\ddot x+\Omega^2 x = \frac{-e^2E_1 E_0}{\gamma ^2m^2c(\omega_0 + \epsilon \Omega)} \cos (\epsilon \Omega t - \Phi_i)
\label{ForcedXequation}
\end{equation}
The solution is the sum of a freely oscillating motion $x^0(t)$ resulting from the injection conditions of the free electron into the standing wave, as given by (\ref{freeXmotion}), and of a forced term $\delta x$ :
\begin{equation}
\delta x (t) = \frac{-e\epsilon E_1 t}{2^{3/2}\gamma m(\omega_0 + \epsilon \Omega)} \sin (\epsilon \Omega t - \Phi_i)
\label{ForcedXmotion}
\end{equation}
It is worth to note that its amplitude almost does not depend on the laser field $E_0$.

We now aim to infer from the electron forced oscillation the work induced by the X-ray field $E_1$ onto the electron velocity along $y$ induced by the laser field.
Let us start by the anti-Stokes case $\epsilon=+1$.

The $y$-motion of the forced electron can be deduced from its $x$-motion as :
\begin{equation}
\delta \ddot y = \frac{-2eE_0}{\gamma m} \sin (k_0x(t))cos(\omega_0 t)
\end{equation}
We assume again that the electron remains close to the bottom of the potential well, so that $sin(k_0 x(t)) = k_0 x^0(t)+k_0 \delta x(t) $; the $y$ velocity can therefore be approximated by the sum of the velocity of the free $y$ motion of Eq. (\ref{freeXmotion}), and a forced velocity $\delta \dot y$:
\begin{equation}
\delta \dot y=\frac{-e^2E_0E_1k_0t}{2^{3/2}\gamma^2 m^2c (\omega_0+\Omega)^2}\cos[(\omega_0+\Omega)t-\Phi_i].
\end{equation}
The average value of the work of the force $-eE_1$ per unit time is therefore :
\begin{equation}
P_{AS} = <\delta \dot y F_1> = \frac{e^3E_0E_1^2 t}{2^{5/2}\gamma^2m^2c (\omega_0+\Omega)^2}
\end{equation}
This power is positive, meaning that the electron gains energy, and conversely that the X-ray wave loses energy. This corresponds necessarily to a damped propagation mode for $E_1$.

If we now turn to the Stokes ($\epsilon =-1$) case, the forced $y$-velocity is :
\begin{equation}
\delta \dot y=\frac{e^2E_0E_1k_0t}{2^{3/2}\gamma^2 m^2c (\omega_0-\Omega)^2}\cos[(\omega_0-\Omega)t-\Phi_i].
\end{equation}
resulting in a negative power transfer :
\begin{equation}
P_{S} = <\delta \dot y F_1> = \frac{-e^3E_0E_1^2 t}{2^{5/2}\gamma^2m^2c (\omega_0-\Omega)^2}
\end{equation}
The Stokes scattered X-ray wave will therefore gain energy from the interaction with the forced part of the electron motion. The increase of $E_1$ will result in an enhanced forced motion $\delta x$ and $\delta y$, which will increase in turn the power transfer to $E_1$. We can therefore expect an exponential amplification of the Stokes wave, that is, to start a stimulated Raman scattering process in the forward direction with respect to the electron beam.

\section{Analysis of the amplification process}
\label{sec:amp}

The analysis of the previous section was purely based on a kinetic, single electron description. We now turn to a macroscopic description, and aim to set the evolution equation along $z$ of an X-ray field $E_1$, coupled to the current density $J_1$ induced by the electron oscillations in the laser field, in conditions where the electrons exhibit bunching in the transverse direction $x$.
We will therefore introduce a mean electron displacement function $\delta x(z,t)$ (illustrated as a red line in Fig. 5(b) ), that will play in the derivation a role very similar to that of the longitudinal bunching factor of Compton Free Electron Laser theory.
From now on, we will focus on the Stokes case.

\subsection{Modal analysis}
\label{ssec:mod}

We start from the exact propagation equation of the X-ray field $E_1$, polarized along the $y$ axis :
\begin{equation}
\frac{\partial^2 E_1}{\partial z^2} + \Delta_\perp E_1 - \frac{1}{c^2}\frac{\partial^2 E_1}{\partial t^2} = \frac{1}{\epsilon_0 c^2} \frac{\partial}{\partial t} J_1
\label{EqPropOrdre2}
\end{equation}

We  split the calculation of Eq. (\ref{EqPropOrdre2}) in three steps :
{\it i/} write down the current density $J_1$ as a function of the mean displacement $\delta x$;
{\it ii/} compute the evolution of $\delta x$ for the electrons, subject to a Lorentz force along $x$ induced the laser B-field and the X-ray field $E_1$ ;
{\it iii/} get back to the  propagation equation (\ref{EqPropOrdre2}), with the newly obtained expression for the current $J_1$.

As the various fields are all slowing evolving in space and time, we will systematically introduce envelope functions, and make use of the slowly varying envelope approximation (SVEA) along the electron motion.

\textit{i/} The current density $J_1(x,z,t)$ can be obtained from :
\begin{equation}
\left( \frac{\partial}{\partial t}+v\frac{\partial}{\partial z}\right)J_1 = \frac{2e^2E_0}{\gamma m} \sum_i \delta (\mathbf{r - r_i}) \sin (k_0x_i) \cos(\omega_0 t),
\label{CurrentDensity}
\end{equation}
where the summation runs over all electrons $i$ in the bunch, and $\mathbf{r}_i$ indicates the  position of electron $i$.
We know from section (\ref{ssec:spa}) that the electron motion along $x$ has  a free and a forced  component, $x_i(t)=x_i^0(t)+\delta x_i(t)$. The forced part $\delta x$ is identical for all electrons in a same slice in the electron bunch, assumed for the time being to be mono-energetic, and in the same potential well.
In contrast, summation over all particles contained in a slice along $z$ brings  the total free motion $x_i^0$ contribution to average out to $0$. This allows us to define a transverse displacement function $\delta x (z,t)$ :
 \begin{equation}
\delta x (z,t) = \frac{1}{N(z,z+dz)} \sum_j{x_j(t)} ,
\label{DefDeltaX}
\end{equation}
 where the summation runs over the all $N(z,z+dz)$  electrons contained in the slice between $z$ and $z+dz$ at time $t$, and in the potential well centered at the origin $x=0$. One may note that the transverse displacement of the next potential well, with respect to its center at $x=\lambda _0 /2$, has the opposite value $-\delta x$;
 however the laser electric field is also dephased by $\pi$, so that the resulting polarizations at $\omega_1$ are in phase for all potential wells. Consideration of $\delta x$ around $x=0$ is therefore well suited to the following derivation.
 In the small angle approximation, we also simplify $\sin (k_0x_i)$ to $k_0 \delta x$.

In parallel, we  reduce the fields $E_1(x,z,t)$ and $J_1(x,z,t)$ to their transverse average values $E_1(z,t)$ and $J_1(z,t)$, and introduce the envelope functions $\tilde E_1$, $\tilde j_1$ and $\delta \tilde x$,
 such as  $E_1= \tilde E_1 \exp{i(k_1z-\omega_1t)} + c.c.$, $J_1= \tilde j_1 \exp{i(k_1z-\omega_1t)}+c.c.$,
and $\delta x=\delta \tilde x \exp i(k_1z-(\omega_1-\omega_0)t)+ c.c.$.
To keep consistent with this transverse field averaging, we neglect the diffraction term of Eq. (\ref{EqPropOrdre2}). The spatial average procedure leads to introduce the electron average number density $\rho$.
With these definitions, the envelopes for current density and transverse displacement are related by :
\begin{equation}
\tilde j_1=\frac{i e^2\rho E_0 k_0}{\gamma m (\omega_0-\Omega)} \delta \tilde x
\label{CurrentVsDisplacement}
\end{equation}

\textit{ii/} The transverse displacement function $\delta x$  follows the eulerian analog of Eq. (\ref{ForcedXequation}) :
\begin{equation}
\left( \frac{\partial}{\partial t}+v\frac{\partial}{\partial z}\right)^2 \delta x + \Omega^2\delta x = \frac{-e}{\gamma m} V_1 \times B_0
\label{eulerForcedXequation}
\end{equation}
where we have introduced the velocity field $V_1$ following from :
\begin{equation}
\left( \frac{\partial}{\partial t}+v\frac{\partial}{\partial z}\right)V_1=\frac{-e\tilde E_1}{\gamma m} \exp{i(k_1z-\omega_1 t)} + c.c.
\end{equation}
so that, with $V_1 =\tilde V_1\exp{i(k_1z-\omega_1t)} + c.c.$ :
\begin{equation}
\tilde V_1 = \frac{-ie\tilde E_1}{\gamma m(\omega_0-\Omega)}.
\end{equation}
The Lorentz force term is therefore
\begin{equation}
-e V_1 \times B_0 = \frac{-e^2E_0\tilde E_1}{\gamma mc(\omega_0-\Omega)}\exp{i(k_1z-(\omega_1-\omega_0) t)} - c.c.
\label{LorentzForceTerm}
\end{equation}
where we have dropped non resonant terms at frequency $\omega_1+\omega_0$ (that would correspond to an excitation at the anti-Stokes frequency), and assumed the magnetic field $B_0$ to be constant in the vicinity of the bottom lines of the potential.
We now apply the SVEA to the first term of equation (\ref{eulerForcedXequation}), resulting in :
\begin{eqnarray}
\left( \frac{\partial}{\partial t}+v\frac{\partial}{\partial z}\right)^2 \delta x
&= \big[ - [\omega_0-\omega_1(1-\frac{v}{c})]^2 \nonumber\\
+2i[\omega_0-\omega_1(1-\frac{v}{c})]&\left(\frac{\partial}{\partial t}+v\frac{\partial}{\partial z}\right) \big]\delta\tilde x\,
e^{i(k_1z-(\omega_1-\omega_0)t)}\nonumber\\ &+ c.c.\nonumber
\label{SveaDeltaX}
\end{eqnarray}
If the frequencies fulfill the condition :
\begin{equation}
\omega_1\left(1-v/c\right)=\omega_0 - \Omega,
\label{ResonanceCondition}
\end{equation}
then the linear term in $\delta\tilde x$  cancels out the restoring force of the harmonic potential in Eq. (\ref{eulerForcedXequation}), so that :
\begin{equation}
2i\Omega \left(\frac{\partial}{\partial t}+v\frac{\partial}{\partial z}\right)\delta\tilde x = \frac{-e^2E_0\tilde E_1}{\gamma^2 m^2c(\omega_0-\Omega)}.
\label{EvolutionDeltaX}
\end{equation}
Note that, while Eq. (\ref{FreqOm1}) was the simple result of a Fourier analysis, Eq. (\ref{ResonanceCondition}) should be interpreted as a resonance condition.

Use of equations (\ref{CurrentVsDisplacement}) and (\ref{EvolutionDeltaX}) allows one to evaluate qualitatively the power gained at resonance by the X-ray field,  as $-j_1.E_1 = -2\textit{Re} (\tilde j_1 \tilde E_1^*)$, where $j_1$ is the current induced by a mean displacement induced over an interval $\delta L$ :
\begin{equation}
-j_1.E_1\simeq  \frac{e^4\rho E_0^2k_0 \delta L }{\gamma^3m^3c v\Omega(\omega_0-\Omega)^2 }.2\tilde E_1.\tilde E_1^* > 0
\end{equation}
We therefore recover the conclusion of the single electron analysis, showing that the Stokes mode exhibits amplification, while the anti-Stokes mode, described simply by replacing $\Omega$ by $-\Omega$, should be absorbed.

\textit{iii/} We eventually come back to the propagation equation (\ref{EqPropOrdre2}), which, under the SVEA, and neglecting diffraction terms, reads :
\begin{equation}
2ik_1 \left( \frac{\partial}{\partial z}+\frac{\partial}{c\partial t}\right)  \tilde E_1 = \frac{-i\omega_1 \tilde j_1}{\epsilon_0c^2}
\label{EqPropSVEA}
\end{equation}

Combining equations (\ref{CurrentVsDisplacement}), (\ref{EvolutionDeltaX}), and (\ref{EqPropSVEA}), and considering the process to be stationary, result in a single propagation equation for the X-ray envelope :
\begin{equation}
\frac{\partial^2\tilde E_1}{\partial z^2} = \frac{e^4\rho k_0 E_0^2}{4\epsilon _0\gamma^3m^3c^2v\Omega (\omega_0-\Omega)^2 } \tilde E_1
\end{equation}
that corresponds to an exponential amplification with a gain of :
\begin{equation}
g = \sqrt {\frac{e^3\rho k_0 E_0}{2^{5/2}\epsilon _0\gamma^2m^2c v(\omega_0-\Omega)^2 }},
\label{gain}
\end{equation}
or, in an approximate simpler way :
\begin{equation}
g = \sqrt {\frac{e^3\rho E_0}{2^{3/2}\epsilon _0 m^2 c^3 \omega_1 }}.
\label{SimplerGain}
\end{equation}

\subsection{Effect of electron velocity mismatch}
\label{ssec:elmis}

The energy dispersion of the incident electron bunch is a major concern for X-ray free electron lasers. In particular, all the simulations on the FEL effect with optical undulators, in the  Compton regime, demonstrate that a remarkable value of mono-energeticity is required, typically of the order of $10^{-4}$ \cite{Bacci} to few $10^{-4}$ for electron energies of few tens of MeV \cite{Petrillo}.
Indeed, in the Compton regime, amplification occurs throughout the laser undulator length only if $\delta \gamma / \gamma < 1/2N$, $N$ being the number of undulator periods over the whole amplification length \cite{RefB_shiozawa}. The Doppler frequency shift is therefore limited to the emission linewidth due to the finite emission time.
This very stringent condition on the electron energy dispersion is obviously one of the major reasons why this optical undulator scheme has not been demonstrated up to now.
How the proposed Raman scheme for a X-ray FEL copes with the electron energy dispersion is therefore a major issue; however, a detailed study of Raman amplification with a  spread of electron energies is beyond the scope of the present study, leading us to restrict ourselves to discuss the spectral broadening induced the electron energy spread, and the amplification regime between a monochromatic X-ray field, and an out-of-resonance electron population.

In general, one has to consider a electron bunch with a distribution of Lorentz factors, with an interval $2\Delta \gamma$ around a central value $\gamma_0$,  characterized by a density distribution $\rho (\gamma)$.
For each velocity component, the deviation $\delta \gamma=\gamma-\gamma_0$ from the central value  results in a shifted X-ray angular frequency $\delta \omega_1$, with $\delta \omega_1 / \omega_1 = 2\delta \gamma / \gamma$. The spontaneous scattering spectrum is therefore bound to exhibit a Doppler broadening of $4\omega_1(\Delta \gamma / \gamma)$.
An outcome of this broadening is the possibility to get spectral overlaps between a Doppler down-shifted emission on a anti-Stokes mode, and a Doppler up-shifted emission on the Stokes mode. Assigning the former to electrons of Lorentz factor $\gamma - \Delta \gamma$ and the latter to those with $\gamma + \Delta \gamma$, the overlap condition reads :
\begin{equation}
\frac{\omega_0-\Omega}{1-v(+\Delta\gamma)/c} = \frac{\omega_0+\Omega}{1-v(-\Delta\gamma)/c}
\end{equation}
Developing to first order results in a simple condition to prevent Stokes / anti-Stokes overlaps :
\begin{equation}
\Delta \gamma /\gamma < \Omega/2\omega_0.
\label{condNoOverlap}
\end{equation}

Another effect due the electron energy spread is that essentially all electrons violate to some degree the resonance condition (\ref{ResonanceCondition}) . We need therefore to evaluate the spectral acceptance of (\ref{ResonanceCondition}). In this aim, we propose to investigate how a monochromatic X-ray field, at the central frequency $\omega_1$,  interacts with
a population of electrons in the bunch, with a Lorentz factor $\gamma '=\gamma +\delta \gamma $, and density $\rho '$.
This simple approach is of course unable to describe the full complexity of the problem, in which each field frequency component is coupled to all electron populations of different velocities, and conversely each electron is coupled to all field frequency components.
It may however give interesting insights on electron - field couplings out of  the resonance condition.

Revisiting the three steps of the gain calculation of section (\ref{ssec:mod}), one may notice that the major effect of the offset in electron kinetic energy, and hence in velocities, is to modify the expression of the derivatives along the movement by adding a term resulting from the velocity change $\delta v\frac{\partial}{\partial z}$.
We will neglect an additional second order effect, namely, the slight change of the resonant oscillation frequency $\Omega$ in the light lattice.

The current density versus transverse displacement function becomes:
\begin{equation}
\tilde j_1=\frac{i e^2\rho E_0 k_0}{\gamma ' m (\omega_0-\Omega-\Delta \omega)} \delta \tilde x,
\label{CurrentVsDisplacementDeltaGamma}
\end{equation}
where we have set $\Delta \omega = k_1.\delta v = \omega_1 \delta \gamma /\beta \gamma ^3$.
The high frequency velocity field is now:
\begin{equation}
\tilde V_1 = \frac{-ie\tilde E_1}{\gamma' m(\omega_0-\Omega-\Delta \omega)},
\end{equation}
so that the new expression of the Lorentz force field is:
\begin{equation}
-e V_1 \times B_0 = \frac{-e^2E_0\tilde E_1}{\gamma' mc(\omega_0-\Omega-\Delta \omega)}e^{i(k_1z-(\omega_1-\omega_0) t)} - c.c.\quad .
\label{LorentzForceTermDeltaGamma}
\end{equation}
The differential equation for transverse displacement function $\delta x$ includes new terms  :
\begin{eqnarray}
-\Delta\omega(\Delta \omega + 2\Omega )\delta\tilde x  +  &2i(\Omega+\Delta \omega) \left(\frac{\partial}{\partial t}+v'\frac{\partial}{\partial z}\right)\delta\tilde x =\nonumber \\
\lefteqn{
\frac{-e^2E_0\tilde E_1}{\gamma'^2 m^2c(\omega_0-\Delta \Omega)}
}
\label{ModifiedEvolutionDeltaX}
\end{eqnarray}
Assuming (\ref{condNoOverlap}) to be valid, let us define $g'$ and $g_I$  as:
\begin{eqnarray}
g' = \left[\frac{e^4\rho' k_0 E_0^2}{4\epsilon _0\gamma'^3m^3c^2v'(\Omega+\Delta\Omega) (\omega_0-\Omega-\Delta \omega)^2 }\right]^{1/2},\nonumber\\
g_I = -\Delta\omega(\Delta \omega + 2\Omega )/[4(\Omega+\Delta \omega)v'].\nonumber
 \end{eqnarray}
The differential equation for the field envelope $\tilde E_1$ becomes :
\begin{equation}
\frac{\partial^2\tilde E_1}{\partial z^2}-2ig_I\frac{\partial\tilde E_1}{\partial z} - g'^2  \tilde E_1 = 0.
\end{equation}
For small values of $\delta \gamma$, $g'\simeq g$, and the reduced discriminant $D=g^2-g_I^2$ of this second order differential equation is positive,
which yields a complex gain coefficient with a positive real value:
\begin{equation}
g(\delta \gamma) = \sqrt {g^2-g_I^2} + i g_I,
\label{gainDeltaGamma}
\end{equation}
where the imaginary part $g_I$ has the dimension of a wave-vector. In these conditions the field continues to exhibit gain, but with reduced values, and the electron population has a new dispersive effect.
When $\delta \gamma$ becomes such that $g_I = g'$, then the discriminant gets negative, and the gain take purely imaginary values, corresponding to oscillating solutions for $\tilde E_1$ and $\delta\tilde x$, of wavevectors $g_I \pm \sqrt{g_I^2 - g^2}$, implying a regular exchange of energy between the field generated at $\omega_1$ and the population of electrons at $\delta \gamma$, and essentially no net transfer between the field and the electrons at the exit of the interaction region.

To first order in $\delta \gamma$, the discriminant vanishes for $\delta\gamma /\gamma = g/k_0$. This defines what can be called an homogeneous spread as the relative width $2\delta \gamma /\gamma$ for which electrons contribute to a gain at $\omega_1$, and an homogeneous spectral width $\Delta \omega_1^H / \omega_1 = 4g/k_0$.
In realistic conditions, the gain length is bound to be much larger than the laser wavelength, implying that the homogeneous width is likely to be smaller than the inhomogeneous Doppler width.
In principle, this narrow homogeneous width should allow stimulated Raman scattering even in conditions of large electron energy spread.
These elementary considerations will have to be revisited however in  more general studies on the effects of electron energy spread.

\subsection{Effect of potential anharmonicity and bunch emittance}
\label{ssec:anharm}
We have so far assumed the electron wiggling to occur very close to the bottom of the ponderomotive potential, so that the harmonic potential approximation could hold. The assumption is valid if the initial transverse transverse velocity of the electrons at the start of the injection process is extremely small, as would result from a very good beam emittance.
However, injection calculations from section \ref{sec:KapitzaDirac}, performed in conditions of the currently best achieved values of beam normalized emittance ($\epsilon _N = 1 \mu m$), show that a number of electrons may also depart from this approximation, and therefore display reduced oscillation frequencies in the ponderomotive potential.
By analogy with usual lasers, we will consider each electron as occupying a "site" given by  its position in phase space, as resulting from the injection process, and corresponding to a unique trajectory $x_0(t)$; the forced transverse motion $\delta x$ follows the equation, extended from Eq. (\ref{ForcedXequation}) :
\begin{equation}
m\gamma\left(\ddot x_0+\delta \ddot  x\right) +\frac{e^2E_0^2}{\gamma m \omega_0^2} \sin\left[2k_0(x_0 + \delta x)\right] = F_L,
\label{ForcedXequationNoApprox}
\end{equation}
where $F_L$ denotes again the Lorentz force; developing to second order with respect to $\delta x$, one obtains :
\begin{equation}
\delta \ddot x +J_0(2k_0x_0)\Omega^2 \delta \ddot x = F_L /\gamma m
\end{equation}
where $x_0$ is the maximum excursion of the electron in the potential well, and $J_0$ is the zeroth-order Bessel function of the first kind. We have neglected here periodic potential terms for $\delta x$,  resulting in a Matthieu-type equation, but bound to average out to zero for many electrons.
The oscillation eigenfrequency is then reduced with respect to the harmonic potential value, as :
\begin{equation}
\Omega ' = \sqrt J_0 (2 k_0 x_0) \Omega,
\label{AnharmFreq}
\end{equation}
The transverse displacement function $\delta \tilde x$ can easily be shown to follow  :
\begin{equation}
\left[\frac{\partial ^2}{\partial z^2}-i\frac{\Delta \Omega^2}{2\Omega v}-\frac{e^4\rho k_0 E_0^2}{4\epsilon_0\gamma^3m^3c^2v\Omega(\omega_0-\Omega)^2}\right]\delta \tilde x = 0 ,
\end{equation}
resulting into a modified gain factor :
\begin{equation}
g'=\sqrt{g^2-\frac{\Omega^2}{v^2}(2k_0x_0)^2}
\label{ReducedGainX0}
\end{equation}
If the electron population is spread over a large distribution in transverse phase space, then the amplification spectrum is broadened following Eq. (\ref{AnharmFreq}), with a reduced gain function depending on the frequency $\Omega '$, given by Eq. (\ref{ReducedGainX0}).
This situation is again typical of an inhomogeneously broadened laser line. The total spectral width of the lasing depends therefore on a combination of Doppler broadening, due to the finite $\delta \gamma/\gamma$ of the electron bunch, and of emittance broadening.
The drawback of a reduced small signal gain is counter-balanced by an important advantage, namely, one can expect the scheme to be  robust with respect to initial spreads in phase space, either longitudinally (energy spread) or transversally (emittance).

\section{Experimental perspectives}
\label{sec:exp}
Several important issues have to be worked out to consider an experimental implementation of this Kapitza-Dirac-Raman X-ray free electron laser.
We will not attempt to address all issues, but only to
 give order-of-magnitude parameters, in order to assess the general experimental feasibility of the proposed scheme.

\subsection{Implementation possibilities for the laser and electron acceleration systems}

We consider laser intensities at focus in the range from $10^{15}$ to $10^{18} W/cm^2$, and laser wavelengths of typically 800 $nm$ or 1.05 $\mu m$. Longer wavelength lasers, such as mid-infrared (resulting eg from an optical parametric chirped pulse amplification process) or far-infrared (CO$_2$ lasers) may be advantageous, but are currently more difficult to implement.
The laser pulse duration should be long enough for the pulse length to be larger that the active region, which corresponds to typical values between few femtoseconds and few hundreds of femtosecond.

Trapping of electrons in the $y$-direction in the active area is an important issue. Several solutions can be considered; one may for instance adopt a 4-wave standing wave geometry, thus providing the same trapping in the $y$ direction as in the $x$ direction. It could offer the advantage of adding a degree of liberty to control the polarization of X-ray light, by controlling the polarization and dephasing of the $y$ lasers.
A second possible solution would be to purposefully shear one beam with respect to the other in the $y$ direction; the standing wave would then be suppressed on both sides, thus creating lateral potential walls of $V_p /4$ (Eq. \ref{VpondStandWaveRelat}). A third possibility could be to irradiate a specially shaped a third beam along $x$, or to alter in a controlled way one of the two twin beams. Several options seem therefore possible, that have to be investigated.

As concerns the electron acceleration setup, one should fully consider the opportunities of the two families of electron accelerators can be considered : conventional RF acceleration, or laser wakefield acceleration.

The major advantage of laser acceleration is to provide extremely short bunches of electrons, with a corresponding very high current density. Moreover, synchronization between the laser-accelerated electron bunch, and the transverse twin lasers, can easily be performed with few femtosecond resolution, if the twin beams are derived from the same laser system, or at least from the same laser oscillator, as the intense laser inducing wakefield acceleration.
Typical values of electron beam currents can reach 10 $kA$ or more, with good emittance values, and very small bunch transverse sizes, of the order of one to few $\mu m$. This scheme suffers from two potential drawbacks : the stability of the electron bunch after the exit of the accelerating plasma, which is the price to pay for such high current densities; and an important value of $\delta E/E$ , whose best measured values are currently in the few percents range.
Drawback $(i)$ can be compensated if one succeeds to get hold of the electron beam in the laser standing wave almost immediately after the exit of the plasma; problem $(ii)$ could be strongly attenuated in the near future, as a number of numerical simulations suggest the possibility to improve mono-energeticity, through an enhanced control of the injection of electrons in the plasma wake. Generally speaking, laser-acceleration of electrons offers extremely promising prospects, especially if the energy dispersion can be reduced experimentally in the per-cent range.

On the other hand, conventional RF acceleration in  a LINAC is a well-known and more mature technology, with a number of existing systems proposing electron bunches up to few tens of MeV, with energy spreads below 1\%, and a normalized emittances down to below 1 $mm.mrad$, especially thanks to the introduction of emittance compensation schemes. Typical peak current values are in the range of 100 $A$; use of magnetic chicanes, such as those set up for the Compton FEL laser projects LCLS and TESLA-XFEL, allows one to reach peak currents up to 3 kA, at the cost of an increased normalized emittance.
An inherent difficulty of conventional RF acceleration is the synchronization issue between the incident electron bunch, and the interference region of the twin laser beams. However, the reliability, and control over conventional LINACs are very good, with the possibility to tune the electron energy, and to control the electron focal position and spot size.
As a result of this alternative, we now present estimates of experimental parameters in both schemes.

\subsection{Prospective implementation parameters}

\begin{table*}
\begin{center}
\caption{Electron bunch, laser and interaction geometry parameters considered}
\label{tab:1}
\begin{tabular}{cccc}
\hline\noalign{\smallskip}
& Laser plasma accelerator & Low energy LINAC & Medium energy LINAC\\
\noalign{\smallskip}\hline\noalign{\smallskip}
Electron energy & 60 MeV & 10 MeV & 155 MeV\\
X-ray photon energy & 43 keV & 1.2 keV & 220 keV\\
Peak current & 25 kA & 2 kA & 100 A\\
Norm. emittance & 1 mm.mrad & 2 mm.mrad & 1 mm.mrad \\
Electron spot size $\sigma _x$ & 1 $\mu$m & 30 $\mu$m & 20 $\mu$m\\
Laser wavelength & 800 nm & 800 nm & 1.05 $\mu$m \\
Laser pulse duration & 30 fs & 30 fs & 400 fs \\
Laser vertical spot size $D$ & 3 $\mu$m & 3 $\mu$m & 5 $\mu$m \\
Laser intensity & 1.4 10$^{18}$ W.cm$^{-2}$ & 2.5 10$^{16}$ W.cm$^{-2}$ & 1. 10$^{16}$ W.cm$^{-2}$\\
Homogeneous gain & 240 cm$^{-1}$ & 4 cm$^{-1}$ & 0.08 cm$^{-1}$\\
Amplification length & 420 $\mu$m & 2.5 cm & 1.1 m\\
Total laser energy $E_L$ & 1.8 J & 1.1 J & 360 J\\
Reference & \cite{malka,davoine} & \cite{rosenzweig} & \cite{SPARC,emit1}\\

\noalign{\smallskip}\hline
\end{tabular}
\end{center}
\end{table*}

Based on the various technological approaches mentioned, we now suggest a few scaling laws and order-of-magnitude parameters for an experimental implementation.
We first reformulate the gain formula, using standard experimental parameters. The beam density $\rho$ is not usually used, but should be deduced from the peak current $I$, and the equivalent electron focal spot $S$:
\begin{equation}
I = e \rho S c,
\end{equation}
where $S=\sigma _x^2 /2$ is the equivalent spot size, if we assume the electron focusing along $x$ and $y$ to be equivalent.

The homogeneous gain formula (\ref{SimplerGain}) becomes :
\begin{equation}
g = \sqrt {\frac{e^2 I E_0}{2^{5/2}\epsilon _0\gamma^2 m^2 c^4 S \omega_0 }}.
\label{GainWithCurrent}
\end{equation}

For the sake of simplicity, we will rely  on this homogeneous gain formula to discuss prospective experimental parameters; such effects as inhomogeneous broadening due to electron dispersion or finite beam emittance, or the transverse bunching of electrons within the standing wave, will be investigated in a full numerical study.

The electron kinetic energy is fixed in a straightforward way by the ratio between the laser and the desired X-ray photon energies, related to first order by $\hbar \omega _1 = 2\gamma ^2 \hbar \omega _0$. The electron technology used to accelerate the electrons then provides fixed values for the emittance $\epsilon _N$ and mono-energeticity $\delta E/E$, summarized in few cases in table \ref{tab:1}.
Two main parameters have to be chosen at that stage : the transverse size $D$ of the active region, and the rms radius $\sigma _x$ of the electron spot size. At saturation, the X-ray output will be optimized if $D\simeq 2 \sigma _x$; however, it may prove useful to concentrate on an active region smaller that the electron beam, in order to enhance the gain by concentrating the available laser power into a small volume.

In laser wakefield acceleration, and direct injection into a light standing wave, $\sigma _X$ is unlikely to be a free parameter; in RF acceleration, there is on the contrary a certain flexibility to choose $\sigma _x$ by playing with the $\beta$  parameter of the focusing magnets. In all cases, we will assume for simplicity that the characteristic sizes are the same in the $x$ and $y$ directions.

From the beam charge, size and duration, one can easily infer the electron current or density, which, coupled to realistic parameters for the dressing twin laser beams, allows one to deduce an order of magnitude of the small signal gain, in the homogeneous limit, and of the total laser energy required to reach a gain.length product of 10 in the electric field, or equivalently of 20 in X-ray intensity.

Table 1 gives the result in the case (i), of an electron bunch resulting from laser wakefield acceleration \cite{davoine,malka} (60 MeV, column 1), (ii) of an electron bunch issued from a state-of-the-art linear accelerator , with either a small (10 MeV, column 2) or medium (155 MeV, column 3) electron kinetic energy. In the first two cases, we make the assumption that the dressing laser is a Titanium-Sapphire system, with a pulse duration of 30 fs; in the last case, we consider typical parameters of a Neodymium-glass laser, with a pulse duration of 400 fs.

The laser intensity is chosen so that the maximum ponderomotive potential of the light lattice is higher that the maximum transverse kinetic energy of the electrons in the bunch, resulting from the normalized emittance  $\epsilon^N$ and the bunch size $\sigma_x$.
The normalized emittance defines the rms  $\sigma _v$ of the transverse velocity distribution :
\begin{equation}
\sigma_v=\epsilon^N/\beta \gamma c \sigma_x ,
\end{equation}
yielding an upper limit for the transverse kinetic energy of :
\begin{equation}
K_\perp = 0.5 \gamma m \sigma_v^2 .
\end{equation}
We suggest a criterium of laser intensity to be defined by a ratio between the maximum ponderomotive potential of Eq. (\ref{VpondStandWaveRelat}), expressed with the usual form $0.9I \lambda^2$, and $K_\perp$:
\begin{equation}
0.9 . (4I) \lambda^2 / \gamma > 5 K_\perp ,
\end{equation}
where the factor of 5 is here largely arbitrary. The factor of 4 originates from the beating between the twin lasers in the interaction region, as shown by Eq. (\ref{VpondStandWaveRelat}).

In the first column, we consider conditions of a laser-plasma accelerator predicted by Davoine \textit{et al.} \cite{davoine}, with an electron bunch accelerated to 60 MeV, a fairly conservative value, and a laser intensity of 1.4 10$^{18}$ W/cm$^2$. The corresponding X-ray energy is as high as 43 keV, beyond the upper limit of X-ray photons expected with Compton Free electron laser.
One sees  that the predicted small signal gain is extremely high, allowing the amplification process to reach saturation over a very short length, which limits the total energy used to values  within the current state of the art in laser technology (Joule class short pulse lasers).
 
In the second column, we  apply the same procedure to the case of a  photo-gun, yielding 10 MeV electrons, assumed to undergo emittance compensation, and beam compression devices, thereby reaching a high peak current of  2 kA, similar to that achieved at higher energies in the TESLA and LCLS projects. While these parameters are obviously very challenging, the corresponding beam brightness of 5 10$^{14}$ A/(m.rad)$^2$ remains well below the maximum value of 3.75 10$^{15}$ A/(m.rad)$^2$ predicted by Rosenzweig \textit{et al.} \cite{rosenzweig}.

Finally, we take in the third column the characteristics of the SPARC system in Frascati \cite{SPARC,emit1}, with state-of-the-art emittance control, but peak current of the order of 100 A. The laser energy required to reach saturation is much higher in this case, but remains in the typical parameters for PetaWatt Nd:glass systems, like the VULCAN laser \cite{vulcan}. It shows however that other options can be considered than ultra-short pulse lasers, that may result in X-ray photon energies reaching the hard X-ray range.

The values obtained in this table, especially for the required laser energies, should be considered merely as order of magnitudes; indeed, our scaling laws are based on a simple theoretical model, that neglects  inhomogeneous broadening and diffraction effects, which will tend to lower the small-signal gain, and on the other hand neglects the increase in electron density in the bottom of the light potential wells, which will have the opposite effect.
While more thorough studies are obviously required, these estimates do raise hope that the Raman X-ray laser scheme could be demonstrated with present day laser technology.

\section{Perspectives and conclusion}
\label{sec:persp}

We have explored the specificities of a novel interaction geometry between a bunch of moderately relativistic electrons, and a standing wave formed by twin high intensity laser beams. We have shown numerically that, in the high intensity regime of Kapitza-Dirac effect, relativistic electrons may get trapped into the minima of the ponderomotive potential, and be guided until the end of the transverse standing wave. 

The electrons tend to oscillate close to the bottom of the potential wells, resulting into a Raman splitting of the scattered radiation in the forward direction. We have shown numerically and analytically that the Stokes component may be coupled back to the transverse electron motion, thereby triggering a stimulated Raman scattering. This can be considered as a new kind of free electron laser effect, in which the electron bunching is no longer longitudinal but transverse.
The scheme seems to display the capability to accept less stringent parameters of bunch
mono-energeticity. This specific robustness may be a key to develop X-ray free electron lasers in the interaction between high intensity lasers, and relativistic electron bunches.

Many aspects of the proposed scheme remain however to be studied, both theoretically and experimentally, in order to ascertain its feasibility and its real potential for applications  : electron injection regime, space charge effects, electron recoil effects, broadening mechanisms, effect of $y$-trapping on the electron dynamics and X-ray wave amplification, Stokes - anti-Stokes couplings, saturation, coherence properties, possibility of X-ray injection, more complex standing wave patterns...
From an experimental point of view, several bottlenecks need to be solved, especially concerning the implementation of the inhomogeneous wave, and the synchronization between the electron bunch and the laser standing wave. The possibility to couple this scheme to setups of laser-plasma wakefield acceleration should  be especially considered.

If a number of positive answers for all these pending physics issues are obtained, and robust implementation schemes are designed, then this scheme may hold the potential to provide compact sources of intense coherent X-ray radiation, with a large number of potential applications in science, from physics to medicine, and technology.

\section{Acknowledgements}
The author would like to acknowledge fruitful discussions with N. Artemiev, V. Malka, and V. Tikhonchuk.

%

%
%

\end{document}